\documentclass{article}
\usepackage{ismir,amsmath,cite,url}
\usepackage{graphicx,import}
\usepackage{color}

\usepackage{booktabs} 
\usepackage[group-separator={,}]{siunitx} 
\usepackage{caption} 
\usepackage[flushleft]{threeparttable} 

\graphicspath{{figs/}}

\title{Modeling Harmony with Skip-grams} 

\multauthor
{David R. W. Sears \hspace{1cm} Andreas Arzt \hspace{1cm} Harald Frostel} { \bfseries{Reinhard Sonnleitner \hspace{1cm} Gerhard Widmer \hspace{1cm}}\\
 Department of Computational Perception, Johannes Kepler University, Linz, Austria\\
{\tt\small david.sears@jku.at}
}
\def\authorname{Sears, Arzt, Frostel, Sonnleitner, Widmer}

\begin{document}

\maketitle
\begin{abstract}
String-based (or \textit{viewpoint}) models of tonal harmony often struggle with data sparsity in pattern discovery and prediction tasks, particularly when modeling composite events like triads and seventh chords, since the number of distinct $n$-note combinations in polyphonic textures is potentially enormous. To address this problem, this study examines the efficacy of \textit{skip-grams} in music research, an alternative viewpoint method developed in corpus linguistics and natural language processing that includes sub-sequences of $n$ events (or \textit{n}-grams) in a frequency distribution if their constituent members occur within a certain number of skips. 

Using a corpus consisting of four datasets of Western classical music in symbolic form, we found that including skip-grams reduces data sparsity in $n$-gram distributions by (1) minimizing the proportion of $n$-grams with negligible counts, and (2) increasing the coverage of contiguous $n$-grams in a test corpus. What is more, skip-grams significantly outperformed contiguous \textit{n}-grams in discovering conventional closing progressions (called \textit{cadences}).
\end{abstract}
\section{Introduction}\label{sec:introduction}

Corpus studies employing string-based (or \textit{viewpoint}) methods in music research often suffer from the \textit{contiguity fallacy}---the assumption that note or chord events on the musical surface depend only on their immediate neighbors. For example, in symbolic music corpora, researchers often divide the corpus into contiguous sequences of \textit{n} events (called \textit{n}-grams) for the purposes of pattern discovery \cite{Conklin:2002}, classification \cite{Conklin:2013}, similarity estimation \cite{Mullensiefen:2009}, and prediction \cite{Pearce:2005}. And yet since much of the world's music is hierarchically organized such that certain events are more stable (or prominent) than others \cite{Bharucha:1983}, \textit{non-contiguous} events often serve as focal points in the sequence \cite{Gjerdingen2014}. As a consequence, the contiguous \textit{n}-gram method yields increasingly sparse distributions as $n$ increases, resulting in the well-known \textit{zero-frequency problem} \cite{Witten:1991}, in which $n$-grams encountered in the test set do not appear in the training set. Perhaps worse, the most highly recurrent temporal patterns in tonal music---melodic formul\ae{}, conventional chord progressions, etc.---are rarely included.

\begin{figure}[t!]
	\hspace{.1em}
	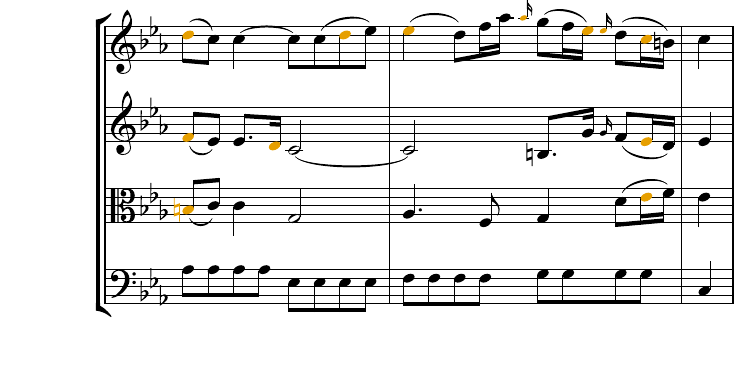
	\caption{Haydn, String Quartet in C minor, Op. 17/4, i, mm. 6--8. Non-chord tones are shown with orange noteheads, and Roman numeral annotations appear below, with the chords of the perfect authentic cadence (PAC) progression embraced by a horizontal square bracket.}
	\label{fig:haydn}
\end{figure}


By way of example, consider the closing measures of the main theme from the first movement of Haydn's string quartet Op. 17, No. 4, shown in \figref{fig:haydn}. The passage culminates in a \textit{perfect authentic cadence}, a syntactic closing formula that features a conventional chord progression (V--I) and a falling upper-voice melody ($\hat{2}$--$\hat{1}$). In the music theory classroom, students are taught to reduce this musical surface to a succession of chord symbols, such as the Roman numeral annotations shown below. Yet despite the ubiquity of this pattern throughout the history of Western tonal music, string-based methods generally fail to retrieve this sequence of chords due to the presence of intervening non-chord tones (shown in orange), a limitation one study has called the \textit{interpolation problem} \cite{Collins:2016}.


To discover the organizational principles underlying tonal harmony using data-driven methods, this study examines the efficacy of \textit{skip-grams} in music research, an alternative viewpoint method developed in corpus linguistics and natural language processing that includes sub-sequences in an \textit{n}-gram distribution if their constituent members occur within a certain number of skips. In language corpora, skip-grams have been shown to reduce data sparsity in \textit{n}-gram distributions \cite{Guthrie:2006}, discover multi-word expressions (or \textit{collocations}) in pattern discovery tasks \cite{Smadja:1993}, and minimize model uncertainty in word prediction tasks \cite{Goodman:2001}. 

Models for the discovery of harmonic progressions in polyphonic corpora typically exclude higher-order sequences (when $n>2$) due to the sparsity of their distributions \cite{Pearce:2004}, so this paper examines the utility of skip-grams for 2-grams, 3-grams, and 4-grams. We begin in Section \ref{sec:representing_chords} by describing the \textit{voice-leading type} (VLT), an optimally reduced chord typology that models every possible combination of note events in the dataset, but that reduces the number of distinct chord types based on music-theoretic principles. Following a formal definition of skip-grams in Section \ref{sec:skipgrams}, Section \ref{sec:experiments} describes the datasets used in the present research and then presents the experimental evaluations, which consider whether skip-grams reduce data sparsity in \textit{n}-gram distributions by (1) minimizing the proportion of rare $n$-grams (i.e., that feature negligible counts), and (2) covering more of the contiguous \textit{n}-grams in a test corpus. We conclude by considering avenues for future research.  

\section{Data-Driven Chord Typologies}\label{sec:representing_chords}

Corpus studies in music research often treat the \textit{note} event as the unit of analysis, examining features like chromatic pitch \cite{Pearce:2004}, melodic interval \cite{Vos:1989}, and chromatic scale degree \cite{Margulis:2008}. Using computational methods to identify \textit{composite} events like triads and seventh chords in complex polyphonic textures is considerably more complex, since the number of distinct $n$-note combinations associated with any of the above-mentioned features is enormous. 


%

\begin{figure}[t!]
	\hspace{.1em}
	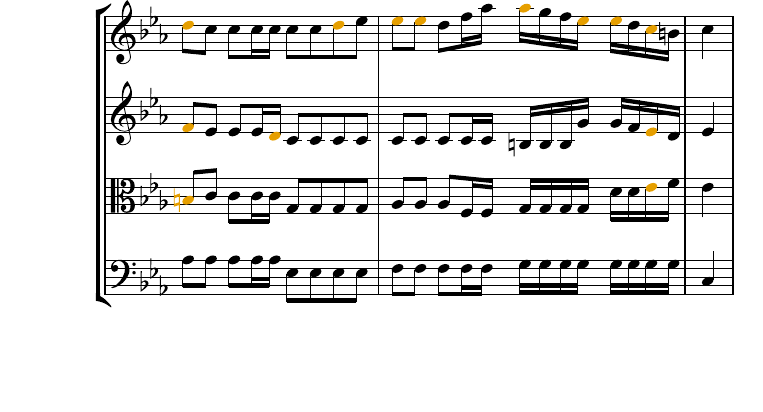
	\caption{Full expansion of Op. 17/4, i, mm. 6--8. Non-chord tones are shown with orange noteheads, and the most representative chord onsets of the PAC progression are annotated with the VLT  scheme.}
	\label{fig:expansion}
\end{figure}

To derive chord progressions from symbolic corpora using data-driven methods, many music analysis software frameworks perform a \textit{full expansion} of the symbolic encoding, which duplicates overlapping note events at every unique onset time.\footnote{In \textit{Humdrum}, this technique is called \textit{ditto} \cite{Huron:1993}, while \textit{Music21} calls it \textit{chordifying} \cite{Cuthbert:2010}.} Shown in \figref{fig:expansion}, expansion results in the identification of 23 unique onset times. Since expansion is less likely to under-partition more complex polyphony compared to other partitioning methods \cite{Conklin:2002}, we adopt this technique for the analyses that follow. 

To reduce the vocabulary of potential chord types, previous studies have represented each chord according to the simultaneous relations between its note-event members (e.g., vertical intervals) \cite{Sears:2016}, the sequential relations between its chord-event neighbors (e.g., melodic intervals) \cite{Conklin:2002}, or some combination of the two \cite{Quinn:2010}. The skip-gram method can model any of these representation schemes, but for the purposes of this study, we have adopted the \textit{voice-leading type} (VLT) representation developed in \cite{Quinn:2010,Quinn:2011}, which produces an optimally reduced chord typology that still models every possible combination of note events in the dataset. The VLT scheme consists of an ordered tuple ($S, I$) for each chord in the sequence, where $S$ is a set of up to three intervals above the bass in semitones modulo the octave, resulting in $13^3$ (or 2197) possible combinations;\footnote{The value of each vertical interval is either undefined (denoted by $\perp$), or represents one of twelve possible interval classes, where 0 denotes a perfect unison or octave, 7 denotes a perfect fifth, and so on.} and $I$ is the melodic interval (again modulo the octave) from the preceding bass note to the present one. 

Because the VLT representation makes no distinction between chord tones and non-chord tones, the syntactic domain of voice-leading types is still very large. To reduce the domain to a more reasonable number, we have excluded pitch class repetitions in $S$ (i.e., voice doublings), and we have allowed permutations. Following \cite{Quinn:2010}, the assumption here is that the precise location and repeated appearance of a given interval are inconsequential to the identity of the chord. By allowing permutations, the major triads $\langle4, 7, 0\rangle$ and $\langle7, 4, 0\rangle$ therefore reduce to $\langle4, 7, \perp\rangle$. Similarly, by eliminating repetitions, the chords $\langle4, 4, 10\rangle$ and $\langle4, 10, 10\rangle$ reduce to $\langle4, 10, \perp\rangle$. This procedure restricts the domain to $233$ unique VLTs when $n=1$ (i.e., when $I$ is undefined). \figref{fig:expansion} presents the VLT encoding for the PAC progression annotated in \figref{fig:haydn}, with the vertical interval classes $S$ provided below each chord onset, and the melodic interval classes $I$ inserted under horizontal angle brackets.

\section{Defining Skip-Grams}\label{sec:skipgrams}

In corpus linguistics, researchers often discover recurrent patterns by dividing the corpus into $n$-grams, and then determining the number of instances (or \textit{tokens}) associated with each unique $n$-gram \textit{type} in the corpus. \textit{N}-grams consisting of one, two, or three events are often called \textit{unigrams}, \textit{bigrams}, and \textit{trigrams}, respectively, while longer \textit{n}-grams are typically represented by the value of \textit{n}. 

\subsection{Contiguous \textit{N}-grams}\label{sec:contiguous}

Each piece $m$ consists of a contiguous sequence of VLTs, so let $k$ represent the length of the sequence in each piece, and let $C$ denote the total number of pieces in the corpus. The number of contiguous \textit{n}-gram tokens in the corpus is 
\begin{equation}\label{eq:1}
\displaystyle\sum_{m=1}^{C} k_m-n+1
\end{equation}
This formula ensures that the total number of tokens is necessarily smaller than the total number of events in the sequence when $n>1$.

\subsection{Non-Contiguous \textit{N}-grams}
The most serious limitation of contiguous \textit{n}-grams is that they offer no alternatives; every event depends only on its immediate neighbors. Without this limitation, the number of associations between events in the sequence necessarily explodes in combinatorial complexity as $n$ and $k$ increase. 

\begin{figure}[t!]
	\centering
	\def\svgwidth{\columnwidth}
	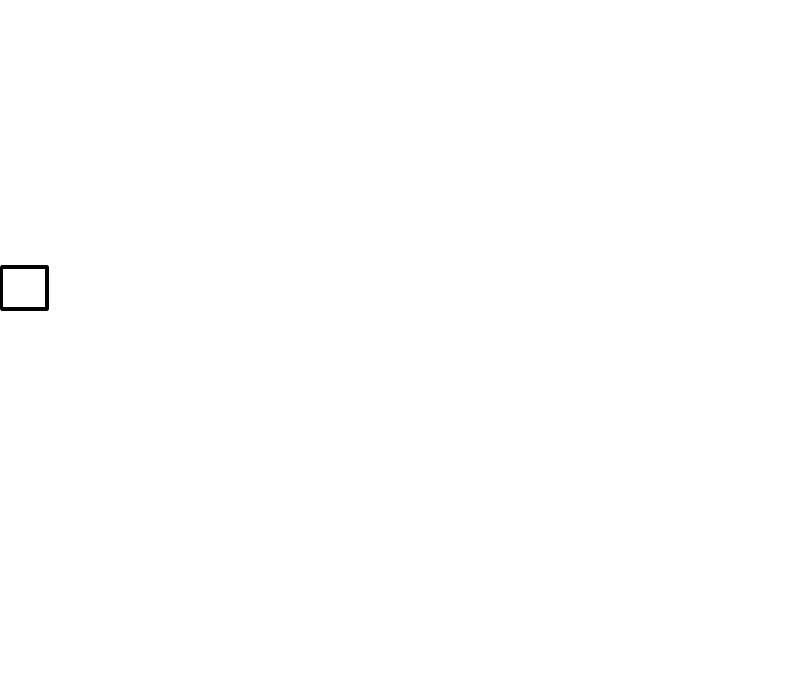
	\caption{Top: A 5-event sequence, with arcs denoting all contiguous (solid) and non-contiguous (dashed) 2-gram tokens. Bottom: All 2-gram tokens, with $t$ indicating the number of skips.}
	\label{fig:non_contiguous}
\end{figure}

The top plot in \figref{fig:non_contiguous} depicts the contiguous and non-contiguous 2-gram tokens for a 5-event sequence with solid and dashed arcs, respectively. According to \eqref{eq:1}, the number of contiguous 2-grams in a 5-event sequence is $k-n+1$, or $4$ tokens. If all possible non-contiguous relations are also included, the number of tokens is given by the combination equation:
\begin{equation}
{k \choose n} = \frac{k!}{n!(k-n)!} = \frac{k(k-1)(k-2) \ldots (k-n+1)}{n!}
\end{equation}

The notation ${k \choose n}$ denotes the number of possible combinations of $n$ events from a sequence of $k$ events. By including the non-contiguous associations, the number of \textit{2}-grams for a 5-event sequence increases to 10. As \textit{n} and \textit{k} increase, the number of patterns can very quickly become unwieldy: a 20-event sequence, for example, contains 190 possible 2-grams, 1140 3-grams, 4845 4-grams, and  15,504 5-grams.

\subsubsection{Fixed-Skip \textit{N}-grams}\label{sec:fixedskip}
To overcome the combinatoric complexity of counting tokens in this way, researchers in natural language processing have limited the investigation to what we will call \textit{fixed-skip} \textit{n}-grams \cite{Guthrie:2006}, which only include \textit{n}-gram tokens if their constituent members occur within a fixed number of skips $t$. Shown in the bottom plot in \figref{fig:non_contiguous}, $ac$ and $bd$ constitute 1-skip tokens (i.e., $t=1$), while $ad$ and $be$ constitute 2-skip tokens. Thus, up to 7 tokens occur when $t=1$, up to 9 occur when $t=2$, and up to 10 occur when $t=3$. 

\subsubsection{Variable-Skip \textit{N}-grams}\label{sec:variableskip}
For natural language texts, the temporal structure of a sequence of linguistic utterances is not clearly defined. Yet for music corpora, temporal characteristics like onset time and duration play an essential role in the realization and reception of musical works. For example, the upper boundary under which listeners can group successive events into temporal sequences is around 2s \cite{Fraisse:1982}. Thus, as an alternative to the fixed-skip method, we also include \textit{variable-skip} \textit{n}-grams, which include \textit{n}-gram tokens if the inter-onset interval(s) (IOI) between their constituent members occur within a specified upper boundary (e.g., 2s). 

\section{Experimental Evaluations}\label{sec:experiments}

This section describes the datasets in the present research and then examines whether the inclusion of skip-grams (1) minimizes the proportion of $n$-gram types with negligible counts, and (2)  covers more of the contiguous \textit{n}-gram tokens in a test corpus. 

\subsection{Datasets \& Pre-Processing}\label{sec:datasets}

\begin{table}[b!]
	\centering
	\begin{threeparttable}
		\begin{tabular}{l S[table-format=3.0] S[table-format=6.0] S[table-format=3.0]}
			\toprule
			Composer (Performer) & \textit{N}$_{\text{pieces}}$ & {\textit{N}$_{\text{chords}}$} & {\textit{N}$_{\text{tokens}>3}$}\\
			\midrule
			Haydn (Kod\'{a}ly) & 50 & 73704 & 0\\
			Mozart (Batik) & 39 & 63418 & 969\\
			Beethoven (Zeilinger) & 30 & 42157 & 910\\
			Chopin (Magaloff) & 156 & 147871 & 3666\\
			\addlinespace[.2cm]
			\multicolumn{1}{r}{\textit{Total}} & 275 & 327150 & 5545\\
			\bottomrule
		\end{tabular}%
		\begin{tablenotes}
			\footnotesize
			\item \textit{Note}. \textit{N}$_{\text{tokens}>3}$ denotes \textit{n}-gram tokens that initially consisted of more than three interval classes.
		\end{tablenotes}
	\end{threeparttable}
	\caption{Datasets and descriptive statistics for the corpus.}
	\label{tab:corpus}%
\end{table}%

Shown in \tabref{tab:corpus}, this study includes four datasets of Western classical music that feature symbolic representations of both the notated score (e.g., metric position, rhythmic duration, pitch, etc.) and a recorded expressive performance (e.g., onset time and duration in seconds, velocity, etc.). Altogether, the corpus totals over 20 hours of music.

The \textbf{Kod\'{a}ly/Haydn} dataset consists of 50 Haydn string quartet movements encoded in MIDI format \cite{Sears:2016}. The data were manually aligned at the downbeat level to recorded performances by the Kod\'{a}ly Quartet, and then the onset time for each chord event in the symbolic representation was estimated using linear interpolation.  
 
The \textbf{Batik/Mozart} dataset consists of 13 complete Mozart piano sonatas encoded in MATCH format \cite{Widmer:2001}. The data were aligned to performances by Roland Batik that were recorded on a B\"{o}sendorfer SE 290 computer-controlled piano, which is equipped with sensors on the keys and hammers to measure the timing and dynamics of each note \cite{Widmer:2003}.

The remaining two datasets were encoded in MusicXML format, and were also aligned to performances that were recorded on a  B\"{o}sendorfer computer-controlled piano. The \textbf{Zeilinger/Beethoven} dataset consists of 9 complete Beethoven piano sonatas performed by Clemens Zeilinger \cite{Flossmann:2010}, while the \textbf{Magaloff/Chopin} dataset consists of 156 Chopin piano works that were performed by Nikita Magaloff \cite{Flossmann:2010,Flossmann:2010b}.


Performing a full expansion on all four datasets produced 327,150 unique onsets from which to derive chords. Unfortunately, some onsets presented more than three vertical interval classes, but since the VLT scheme only permits up to three interval classes $S$ above the bass, it was necessary to replace these chords. Each onset containing more than three distinct vertical interval classes was replaced either with (1) the closest maximal subset estimated from the immediate surrounding context (i.e., $\pm 5$ chords); (2) the most common maximal subset estimated from the entire piece; or finally (3) the most common maximal subset estimated from all pieces in the corpus.

\subsection{Reducing Sparsity}\label{sec:distributions}

\begin{table}[b!]
	\centering
	\begin{threeparttable}
		\renewcommand{\TPTminimum}{\columnwidth} 
		\makebox[\columnwidth]{
			\begin{tabular}{r S[table-format=8.0] S[table-format=9.0]}
				\toprule
				\multicolumn{1}{l}{Skip} & {$N_{\text{types}}$} & {$N_{\text{tokens}}$}\\
				\midrule
				\multicolumn{3}{l}{\textit{No Skip}} \\
				&        135331  &          326034  \\
				\multicolumn{3}{l}{\textit{Fixed -- Skip boundary} (\#)} \\
				1     &        850222  &       2604972 \\
				2     &     2364840  &       8780643 \\
				3     &     4765289  &     20786976 \\
				4     &     8207123  &     40548000 \\
				\multicolumn{3}{l}{\textit{Variable -- IOI}\tnote{a} \hspace{.15em} \textit{boundary} (\textit{s})} \\
				0.5   &     2213148  &     10150852  \\
				1     &   12498736  &     90278381  \\
				1.5   &   31591468  &   306289766  \\
				2     &   59147107  &   718717231  \\
				\bottomrule
			\end{tabular}%
		}
		\begin{tablenotes}
			\footnotesize
			\item[a] IOI denotes the maximum permitted inter-onset interval in seconds between adjacent members of each \textit{n}-gram.
		\end{tablenotes}
	\end{threeparttable}
	\caption{Counts associated with 4-gram types and tokens using both fixed and variable skips.}
	\label{tab:fourgram_counts}%
\end{table}%

In natural language corpora, \textit{n}-gram distributions of individual words ($n=1$) and multi-word expressions ($n<5$) demonstrate a power-law relationship between frequency and rank, with the most frequent (i.e., top-ranked) types accounting for the majority of the tokens in the distribution \cite{Williams:2015}. In music corpora, however, this relationship becomes increasingly linear as $n$ increases due to the greater proportion of types featuring negligible counts. Such rare $n$-grams are thus more difficult to retrieve and model in discovery and prediction tasks, so this section examines whether the inclusion of skip-grams minimizes the proportion of rare $n$-grams in chord distributions.

\subsubsection{Methods}\label{sec:distributions_methods}

Contiguous \textit{n}-gram distributions were calculated from $n=1$ to $n=7$, along with 4-grams that include the following skip levels: \textit{Fixed} -- up to 1, 2, 3, or 4 skips; \textit{Variable} -- all possible skips occurring within a maximum IOI of .5, 1, 1.5, or 2s.

\subsubsection{Results}\label{sec:distributions_results}

\tabref{tab:fourgram_counts} presents the counts for 4-gram types and tokens with both fixed and variable skips. As expected, including skips of either type significantly increased the number of types and tokens. When skips were not included, the corpus produced over 300 thousand tokens, but this number increased to over 40 million tokens for skip-grams including up to 4 skips, or over 700 million tokens for skip-grams including all skips occurring within an IOI of 2s.  

\begin{figure}[b!]
	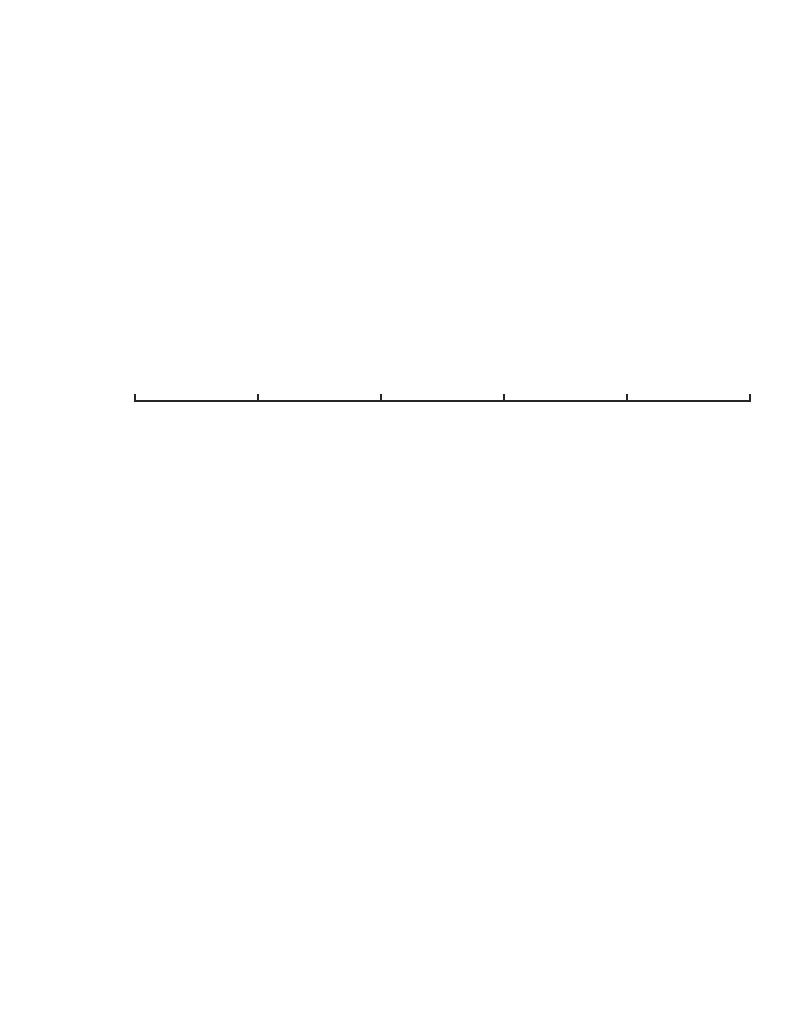
	\caption{Cumulative probability distributions for (top) contiguous \textit{n}-gram types, with types appearing to the right of each marker featuring only one token in the corpus; and (bottom) 4-gram types featuring no skips, up to four skips, or all skips occurring within an IOI of 2s.}
	\label{fig:cum_distributions}
\end{figure}

To visualize the increasing impact of data sparsity on the \textit{n}-gram distribution as $n$ increases, the top plot in \figref{fig:cum_distributions} presents the cumulative probability distributions for contiguous \textit{n}-gram types from $n=1$ to $n=7$. Types appearing to the right of each marker feature only one token in the corpus. When $n$ is small, the distributions loosely conform to the family of power laws used in linguistics to describe the frequency-of-occurrence of words in language corpora, where a small proportion of types account for most of the encountered tokens. When $n$ increases, however, the proportion of types featuring negligible counts also increases, resulting in increasingly uniform distributions. 

Shown in the bottom plot in \figref{fig:cum_distributions}, the power-law relationship returns in the 4-gram distributions when skips are included. What is more, the proportion of types featuring negligible counts also decreases, thereby minimizing the potential for data sparsity in the VLT distribution.


\subsection{Increasing Coverage}\label{sec:coverage}

This section examines whether the inclusion of skip-gram types during training covers more of the contiguous \textit{n}-gram tokens in a test corpus.

\subsubsection{Methods}\label{sec:coverage_methods}

2-gram, 3-gram, and 4-gram distributions were calculated for the following skip levels: \textit{Fixed} -- no skip, or up to 1, 2, 3, or 4 skips; \textit{Variable} -- no skip, or all possible skips occurring within an IOI of .5, 1, 1.5, or 2s. To evaluate skip-gram coverage, we employed 10-fold cross-validation stratified by composer \cite{Dietterich:1998}, using the proportion of contiguous \textit{n}-gram types in the test set that appeared in the training set as a measure of performance. To create folds containing the same number of compositions \textit{and} chords, we computed the mean number of chords that should appear in each fold $m$, and then selected the fold indices for which each fold (1) contained an approximately equal number of compositions, and (2) contained a total number of chords that was $\pm1$\% of $m$. 

\subsubsection{Analysis}\label{sec:coverage_analysis}

To examine the potential increase in coverage at each successive (fixed or variable) skip, we calculated a planned comparison statistic that does not assume equal variances, called the Welch $t$ test.\footnote{In hypothesis testing, planned comparisons typically follow an omnibus statistic like the $F$ ratio, which indicates whether the differences between the means of a given factor  are significant. In this case, the Welch $F$ test was significant for every model, so we forgo reporting those statistics here, and instead simply report the planned comparisons, which indicate whether coverage \textit{increased} significantly as the number of skips (or the size of the temporal boundary) increased.} The mean of each skip was compared to the mean of the previous skip using backward-difference coding (e.g., \textit{Fixed}: 2 skips vs. 1 skip, 3 skips vs. 2 skips, etc.). To minimize the risk of committing a Type I error, each comparison was corrected with Bonferroni adjustment, which divides the significance criterion by the number of planned comparisons.

\subsubsection{Results}\label{sec:coverage_results}


\begin{table*}[t!]
	\centering
	\begin{threeparttable}
		\renewcommand{\TPTminimum}{\textwidth} 
		\makebox[\textwidth]{
			\begin{tabular}{rrlcc S[table-format=<0.3,add-integer-zero = false] cc S[table-format=2.3] S[table-format=<0.3,add-integer-zero = false] ccS[table-format=2.3] S[table-format=<0.3,add-integer-zero = false]}
				\toprule
				&       &       & \multicolumn{3}{c}{\textbf{2-grams}} &       & \multicolumn{3}{c}{\textbf{3-grams}} &       & \multicolumn{3}{c}{\textbf{4-grams}} \\
				\multicolumn{2}{l}{Skip} &       & {$M_{\text{coverage}}$} & \textit{t} & {\textit{p}} &       & {$M_{\text{coverage}}$} & {\textit{t}} & {\textit{p}} &       & {$M_{\text{coverage}}$} & {\textit{t}} & {\textit{p}} \\ \cmidrule{1-2}\cmidrule{4-6}\cmidrule{8-10}\cmidrule{12-14}    
				\multicolumn{14}{l}{\textit{No Skip}} \\
				&       &       & .959  &       &       &       & .707  &       &       &       & .365  &       &  \\
				\multicolumn{14}{l}{\textit{Fixed -- Skip boundary} (\#)} \\
				& 1     &       & .976  & 7.144 & <.001 &       & .813  & 9.726 & <.001 &       & .529  & 10.963 & <.001 \\
				& 2     &       & \textbf{.983} & 4.000 & .003  &       & .859  & 5.518 & <.001 &       & .618  & 6.023 & <.001 \\
				& 3     &       & .986  & 2.529 & .085  &       & .884  & 3.620 & .008  &       & .672  & 3.948 & .003 \\
				& 4     &       & .988  & 1.848 & .327  &       & \textbf{.901} & 2.814 & .046  &       & \textbf{.711} & 3.063 & .027 \\
				\multicolumn{14}{l}{\textit{Variable -- IOI boundary} (\textit{s})} \\
				& 0.5   &       & .979  & 8.439 & <.001 &       & .837  & 12.744 & <.001 &       & .595  & 15.795 & <.001 \\
				& 1     &       & .988  & 6.598 & <.001 &       & .904  & 10.132 & <.001 &       & .727  & 9.786 & <.001 \\
				& 1.5   &       & \textbf{.992} & 3.647 & .010  &       & .929  & 5.313 & <.001 &       & .788  & 5.266 & <.001 \\
				& 2     &       & .993  & 2.311 & .132  &       & \textbf{.943} & 3.564 & .009  &       & \textbf{.824} & 3.808 & .005 \\
				\bottomrule
			\end{tabular}%
		}
		\begin{tablenotes}
			\footnotesize
			\item 
		\end{tablenotes}
	\end{threeparttable}
	\caption{Mean coverage estimates and planned comparisons for 2-gram, 3-gram, and 4-gram tokens using either fixed or variable skips.}
	\label{tab:coverage_stats}%
\end{table*}%
 
\begin{figure}[t!]
 	\centering
 	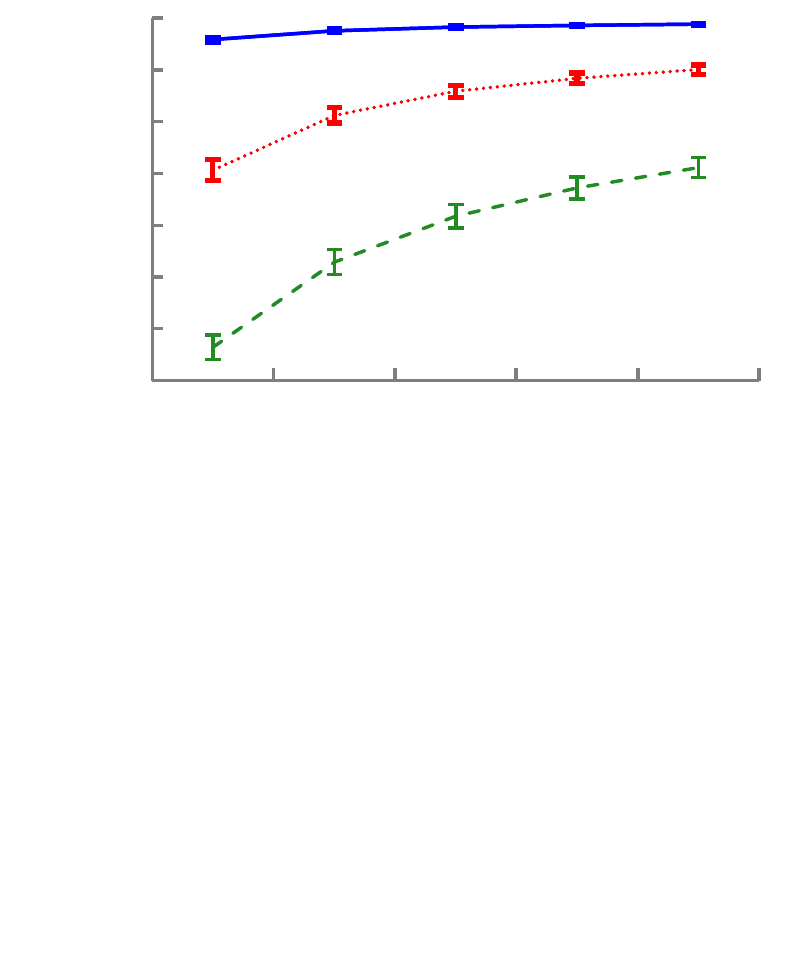
 	\caption{Line plots of the mean proportion of \textit{n}-gram tokens from the test that were covered during training using either fixed (top) or variable (bottom) skips. Whiskers represent the 95\% confidence interval (CI) around the mean.}
 	\label{fig:coverage_lineplots}
 \end{figure}

 \figref{fig:coverage_lineplots} displays line plots of the mean proportion of contiguous \textit{n}-gram tokens from the test that appeared during training using either fixed or variable skips. \tabref{tab:coverage_stats} provides the mean coverage estimates and planned comparisons. For 2-grams, on average the contiguous types covered nearly 96\% of the tokens in the test set. When skips were included, this estimate improved significantly to 98.3\% of the tokens for up to two fixed skips, or up to 99.2\% percent of the tokens for all skips occurring within an IOI of 1.5 s. 
 
 As $n$ increased, the proportion of tokens that appeared during training using contiguous \textit{n}-grams decreased substantially. For 3-grams, the contiguous types only covered 70.7\% of the tokens on average. This estimate improved dramatically when either fixed or variable skips were included, however. For the fixed-skip factor, including up to four skips during training covered an additional 20\% of the tokens during test, resulting in a mean coverage estimate of over 90\%. In the variable-skip condition, this estimate further improved to 94.3\% when all skips occurring within an IOI of 2s were included. Finally, for 4-grams, the contiguous types covered just 36.5\% of the tokens, but this estimate improved to 71.1\% in the fixed-skip condition, and to 82.4\% in the variable-skip condition.

\section{Summary and Conclusion}\label{sec:conclusions}

To reduce data sparsity in \textit{n}-gram distributions of tonal harmony, this study examined the efficacy of skip-grams, an alternative viewpoint method that includes sub-sequences in an \textit{n}-gram distribution if their constituent members occur within a certain number of skips (\textit{fixed}), or a specified temporal boundary (\textit{variable}). To that end, we compiled four datasets of Western classical music that feature symbolic representations of the notated score. Our findings demonstrate that the inclusion of skip-grams reduces sparsity in higher-order \textit{n}-gram distributions by (1) minimizing the proportion of $n$-grams with negligible counts, thus recovering the power-law relationship between frequency and rank when $n<5$ that was previously lost in the corresponding contiguous distributions, and (2) increasing the coverage of the contiguous \textit{n}-grams in a test set, thereby mitigating the severity of the zero-frequency problem.

\begin{table}[b!]
	\centering
	\begin{threeparttable}
		\renewcommand{\TPTminimum}{\columnwidth} 
		\makebox[\columnwidth]{
			\begin{tabular}{r S[table-format=2.0] S[table-format=2.0]}
				\toprule
				\multicolumn{1}{l}{Skip} & {I$^6$-ii$^6$-V$^7$-I} & {ii$^6$-``I$^6_4$''-V$^7$-I}\\
				\midrule
				\multicolumn{3}{l}{\textit{No Skip}} \\
				&        0  &          7  \\
				\multicolumn{3}{l}{\textit{Fixed -- Skip boundary} (\#)} \\
				1     &        3  &       16 \\
				2     &     10  &       36 \\
				3     &     13  &     50 \\
				4     &     15  &     63 \\
				\multicolumn{3}{l}{\textit{Variable -- IOI}\tnote{a} \hspace{.15em} \textit{boundary} (\textit{s})} \\
				0.5   &     5  &     8  \\
				1     &   10  &     33  \\
				1.5   &   21  &   51  \\
				2     &   32  &   77  \\
				\bottomrule
			\end{tabular}%
		}
		\begin{tablenotes}
			\footnotesize
			\item \textit{Note}. VLT encodings for these progressions appear in the major and minor mode, and feature the pre-dominant and dominant harmonies both with and without the seventh (e.g., ii$^6$ and ii$^6_5$).
		\end{tablenotes}
	\end{threeparttable}
	\caption{Number of pieces containing semplice or composta four-chord progressions using both fixed and variable skips.}
	\label{tab:PAC_counts}%
\end{table}%

In our view, this approach would directly benefit tasks related to pattern discovery and prediction, since recurrent temporal patterns rarely appear on the musical surface, thereby forcing \textit{n}-gram models to either exclude higher-order \textit{n}-grams (e.g., where $n>2$) due to the sparsity of the distributions, or calculate \textit{escape probabilities} to accommodate patterns that do not appear (contiguously) in the training set \cite{Cleary:1984}. Consider, for example, the two four-chord cadential progressions in \tabref{tab:PAC_counts}: the \textit{semplice} cadence, which features a dominant-to-tonic progression in root position (e.g., I$^6$-ii$^6$-V$^7$-I); and the \textit{composta} cadence, which also features a six-four suspension above the cadential dominant (e.g., ii$^6$-``I$^6_4$''-V$^7$-I). These cadences are ubiquitous in music of the classical style, and yet the VLT configurations representing these progressions rarely appear on the surface; the semplice cadence \textit{never} appears contiguously, while the composta cadence is featured in just seven pieces. When skips are included, however, the two progressions appear in 32 and 77 of the 245 pieces in the corpus, respectively.

Due to the combinatoric complexity of the task, one limitation of the skip-gram method is that execution times become unfeasible beyond certain values of $n$ and $t$. Nevertheless, if the organizational principles underlying hierarchical stimulus domains like natural language or polyphonic music reflect limitations of human auditory processing, it seems reasonable to impose similar restrictions on the sorts of contiguous and non-contiguous relations the skip-gram method should model. Given the restrictions imposed in this study, retrieving all 4-gram tokens from a sequence of 1,000 chords using commodity hardware produced runtimes of less than 100ms in the largest fixed-skip condition ($t=4$ skips), and less than 3s in the largest variable-skip condition ($t = 2$s), proving skip-gram modeling is entirely attainable in a research setting. 

Of course, counting all possible skip-grams in this way assumes no a priori knowledge about the sorts of non-contiguous relations analysts might hope to discover. For example, collocation extraction algorithms in the NLP community typically exclude infrequent \textit{n}-grams, or use parts-of-speech tags to privilege syntactically meaningful utterances \cite{Smadja:1993}. Music researchers could adopt similar methods by excluding (or weighting) each \textit{n}-gram by the temporal proximity or periodicity of its members \cite{Sears:2016}, or privileging patterns that appear in strong metric positions or feature changes of harmony. Together with the skip-gram method, these techniques could usher in a new suite of inductive, data-driven tools for the discovery of musical organization.

\section{Acknowledgments}\label{sec:acknowledgements}

This research is supported by the European Research Council (ERC) under the EUs Horizon 2020 Framework Programme (ERC Grant Agreement number 670035, project ``Con Espressione'').

\bibliography{ISMIR2017}

%
%
%
%

\end{document}